\documentclass{aa}
\usepackage{psfig,graphicx,natbib}
\newcommand{\Msun}{\mbox{\rm M$_{\odot}$}}
\begin{document}
   \title{Early galaxy evolution from deep wide field star counts}

   \subtitle{II. First estimate of the thick disc mass function}

   \author{C. Reyl\'e
          \inst{1}
          \and
          A.C. Robin \inst{1}
          }

   \offprints{C. Reyl\'e}

   \institute{CNRS UMR6091, Observatoire de Besan{\c c}on, BP1615, 
    F-25010 Besan{\c c}on Cedex, France\\
	      \thanks{Partly based on observations made at CFHT, and at the
European Southern Observatory}
             }

   \date{Received 11 April 2001 / Accepted 23 April 2001}

   \titlerunning{The thick disc mass function}

   \abstract{Star counts at high and intermediate galactic latitudes, in the
visible and the near infrared, 
are used to determine the density law and the initial mass function of the
thick disc population. The combination of shallow fields dominated by stars at 
the turnoff with deep fields allows the determination of the thick
disc mass function in the mass range 0.2-0.8\Msun.
Star counts are compared with simulations of a synthesis population model.
The fit is based on a maximum likelihood criterion. The best fit model gives
a scale height of 800 pc, a scale length of 2500 pc and a local density of 
10$^{-3}$ stars pc$^{-3}$ or 7.1 10$^{-4}$ \Msun pc$^{-3}$
for Mv $\leq$ 8.  
The IMF is found to follow a 
power law dN/dm $\propto$ m$^{-0.5}$. This is the first determination of the 
thick disc mass function.
   }

   \maketitle
%

\section{Introduction}

Solving the problem of the origin of the thick disc of the Milky Way depends
on accurate determination of its present characteristics. Overall
analysis of density, kinematics and chemical properties help the understanding 
of the physical processes involved.
A first step was obtained with the finding that the thick disc
was probably formed by a merging event on the thin disc early in the early age
of the Milky Way \citep{1993MNRAS.262..350S,Robin1996A&A...305..125R}.
This hypothesis was motivated by the kinematic findings (no gradient, no discontinuity 
between thin disc and thick disc in rotation and velocity dispersion) and 
abundances, in particular the [O/Fe] and [Mg/Fe] ratios. These ratios implies 
a sudden decrease in star formation rate between the thick disc and thin disc 
formation, lasting for at least 1 Gyr but not more than 3 Gyr 
\citep{Gratton2000A&A}. More work is required to measure an accurate 
density law and to obtain a detailed description of the stellar population, such 
as the initial mass function and age.
 
The thick disc density law can reasonably be modeled by a double exponential, 
or a density law close to a sech$^2$. Star counts are presently unable to
distinguish between these hypotheses. Even the 
determination of the scale height and the local density causes difficulties due
to a slight degeneracy between these two parameters, as shown in various 
published results. Different analyses have resulted in either
high scale height and small local density (for example, 
\citet{Reid1993ApJ...409..635R}: 1400 pc and local density of 2\% of the disc) 
or small
scale height and higher local density (\citet{Robin1996A&A...305..125R}: 760 
pc and density of 5.6\%, \citet{1999A&A...348...98B}: 910 pc and 5.9\%), with 
several intermediate results.
The number of observed stars is derived by
the integral of the density law over the line of sight. For an exponential 
density law, the mean distance
of the stars in a complete sample is roughly twice the scale height.
The thick disc dominates star counts at distances between 2 and 5
kpc over the galactic plane. However, photometric counts are not accurate enough
to estimate the distances of stars at the turnoff with an accuracy of even
a factor of two. Moreover no accurate determination of the local density
has ever directly been done, even with Hipparcos, because of the small 
proportion of the thick disc locally with regard to the thin disc.

The determination of the initial mass function (IMF) of the thick disc is 
an important issue in the controversy about the universality of the IMF.
\citet{1998simf.conf..201S} and \citet{2001MNRAS.322..231K} find
that it flattens at masses below 0.5\Msun, either in our Galaxy or in the LMC.
Whereas Scalo points out the difficulty measuring an IMF and argues that 
the uncertainties on the determination could be of the order of the apparent 
variations of the IMF, Kroupa analyzes in detail the
variations of the IMF slope and concludes that star formation 
in higher metallicity environments appears to produce relatively more low-mass
stars, i.e. a steeper slope at low masses.
If confirmed, one should expect
the thick disc to have a shallower IMF slope than the thin disc on average.

Until now no direct measurement of the thick disc IMF has been done.
All along the main
sequence, the thick disc population is easily distinguishable from the 
disc and the
halo using a good temperature indicator like the V-I index, 
because at a given magnitude below the turnoff of the halo ($\simeq$ V $>$ 17),
the blue side is dominated by the halo and the red side by
the disc, with the thick disc in between. At high latitudes, thick disc stars become a 
sizeable population at magnitude about 14-15 in V in wide field star counts when 
a significant proportion
of turnoff stars are detectable on the blue side of the colour distribution. 
One can reach the peak of the luminosity function (at about M$_V$=11) at 
magnitude
21 for stars at 1 kpc. By combining shallow star counts dominated by
turnoff stars with deep photometry one should be able to compute the IMF slope
on a large mass range, between the turnoff at about m = 0.8 \Msun and the 
maximum near m = 0.2 \Msun. 

In this paper we address the problem of the thick disc density law together
with its IMF. The two problems cannot be treated separately from
star count analysis. We use a large set of stellar samples 
(described in section~\ref{sec2}), in the
visible and the near infrared, at shallow and deep magnitudes, to investigate
the thick disc luminosity function, the local density and its scale height
and scale length. This analysis has been feasible using a coherent model of
population synthesis which takes into account the various photometric systems
of the data, different basic hypotheses
on the parameters to test, and allows us to disentangle the bias effects in star
count samples (section~\ref{sec3}). A maximum likelihood test is used to estimate
the thick disc parameters (section~\ref{sec4}). Results are given in section~\ref{sec5} 
and discussed in section~\ref{sec6}.


\section{Selected data sets}
\label{sec2}
\begin{figure*}
\centering
\includegraphics[angle=0,width=15cm]{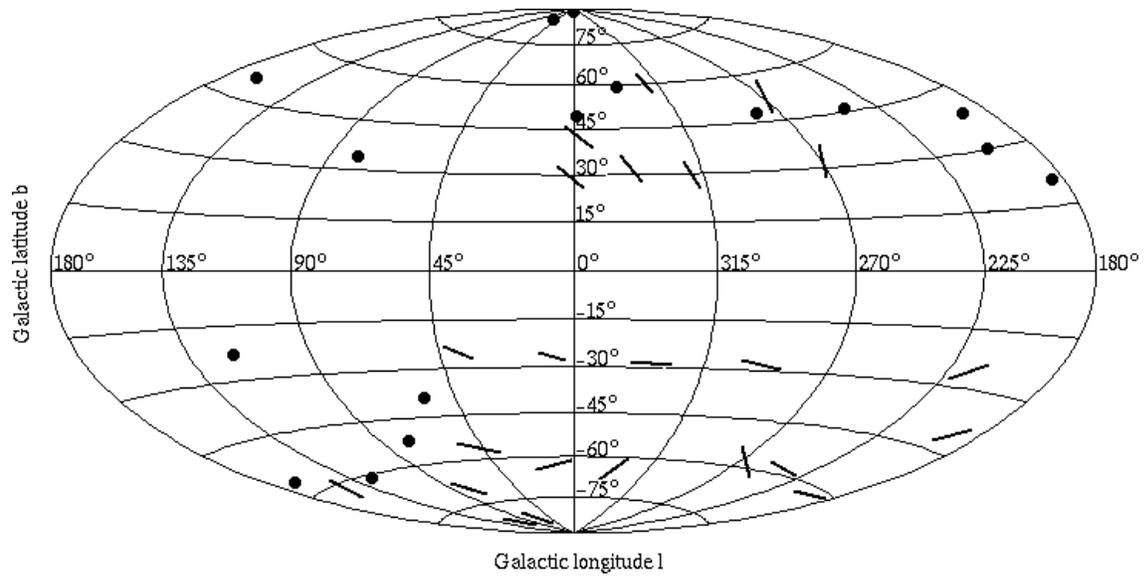}
   \caption{Galactic coordinates of the selected fields. Characteristics
of the fields plotted with dots are given in table~1. The lines 
represent parts of DENIS strips.}
   \label{fig1}
\end{figure*}

Data sets at medium and high galactic latitudes have been selected. They
combine shallow and deep star counts, in the visible
and the near infrared. The main characteristics of visible data are summarized 
in table~1. The dots in figure~\ref{fig1} give their distribution
in galactic coordinates. The shallow star counts are three wide fields 
from \citet{Ojha1996,1999A&A...351..945O}, 
one field from \citet{Chiu80b} close to the galactic 
anticentre and one patch from the CNOC2 Field Galaxy Redshift Survey 
\citep{Yee2000ApJS}. The deep fields are 
one field from \citet{Borra86}, one field from \citet{Reid1993ApJ...409..635R} 
close to the North Galactic Pole, six fields from the DMS 
\citep{Hall1996ApJS,Osmer1998ApJS} used to
search for quasars, two fields from \citet{Bouvier1998A&A} dedicated to the study of Pleiades
and Praesepe clusters, and another field near the North Galactic Pole (SA57)
obtained at the CFHT by Cr\'ez\'e et al. (in preparation). 
This field is the deepest one. It is complete and free from significant 
galaxy contamination up to V=24.

We also used near infrared data in I and J bands from the Deep Near Infrared 
Survey of the Southern Sky: DENIS 
\citep{Epchtein1997Msngr..87...27E,1999A&A...349..236E} reduced at the Paris Data
Analysis Center.  
DENIS fields, named strips, are 12' in right ascension and 30$^\circ$ in 
declination. We have selected 26 strips distributed on the 
sky at latitudes greater than 30$^\circ$, as shown by the lines in 
figure~\ref{fig1}. 
At lower latitudes, the disc population becomes dominant compared to the thick 
disc population. Selected samples are portions of strips over which the 
density gradient in declination is negligible. They cover 1 to 3 square 
degrees and are complete up to I=17.5 or I=18, depending on the strip.

The absolute visual magnitude of thick disc stars in the selected samples
covers a wide range, from M$_V \simeq$ 4 up to 13. It peaks
at M$_V \simeq$ 4-5 in the DENIS fields, M$_V \simeq$ 6-7 in the shallow 
fields, M$_V \simeq$ 9-10 in the deep fields and M$_V \simeq$ 12 in SA57.


\section{The model of population synthesis}
\label{sec3}

We have used a revised version of the Besan\c{c}on model of 
population synthesis.
Previous versions were described in 
\citet{Bienayme1987a,Bienayme1987b,Haywood1997A&A...320..440H,Robin2000A&A}.

The model is based on a semi-empirical approach, where physical constraints
and current knowledge of the formation and evolution scenario of the
Galaxy are used as a first approximation for the population synthesis. 
The model involves 4 populations (disc, thick disc, halo and bulge) 
each deserving a specific treatment. 
The bulge population, which is irrelevant for this analysis, will be 
described elsewhere.

\subsection{The disc population}

A standard evolution model is used to produce the disc population,
based on a set of usual parameters : an initial mass function (IMF), a star
formation rate (SFR) and a set of evolutionary tracks  
\citep[see][and references therein]{Haywood1997A&A...320..440H}. 
The disc population is assumed to evolve over 10 Gyr. 
A set of IMF slopes and SFRs are tentatively assumed and tested against
star counts. 

A revised IMF has been used in the present analysis, tuned with the most recent
Hipparcos results: the age-velocity dispersion relation is from 
\citet{Gomez1997ESASP}, the local luminosity function is from 
\citet{Jahreiss1997ESASP}
and an IMF power law dN/dm $\propto$ m$^{-\alpha}$ is adjusted to it, giving 
a slope $\alpha$ = 1.6 in the low 
mass range 0.08-0.5\Msun, in good agreement with \citet{Mera96b}: $\alpha$ = 
2$\pm$0.5 and \citet{Kroupa2000AGM}: $\alpha$ = 1.3$\pm$0.5.
The scale height has been self-consistently computed using the potential
obtained from the constraints on the local dynamical mass from 
\citet{Creze1998A&A}. This aspect will be described elsewhere (Reyl\'e et al., 
in preparation).

The evolutionary model fixes the distribution of stars in the space 
of intrinsic
parameters : effective temperature, gravity, absolute
magnitude, mass and age. These parameters are converted into
colours in various systems through stellar atmosphere models 
corrected to fit empirical data 
\citep{Lejeune1997A&AS..125..229L,Lejeune1998A&AS..130...65L}. While some errors
still remain in the resulting colours for some spectral types, the overall
agreement is good in the major part of the HR diagram. For low mass stars in
the near infrared, synthetic colours from \citet{1998A&A...337..403B} have been 
used.

Since the \citet{Haywood1997A&A...320..440H} model is based on evolutionary 
tracks at solar metallicities, inverse blanketing corrections are introduced 
to give to the disc a metallicity distribution in agreement with the
\citet{Twarog80} age/metallicity distribution (mean and dispersion about the 
mean).

\subsection{The thick disc population}

In the population synthesis process,
the thick disc population is modeled as originating from a single epoch of
star formation. We use \citet{Bergbush92} oxygen enhanced evolutionary 
tracks. No strong constraint currently exists on the thick disc age. We 
assume an age of 14~Gyr. An age of 11~Gyr, which is slightly older than the 
disc and younger than the halo, does not give significantly different results.

The thick disc metallicity can be chosen between $-0.4$ and $-1.5$~dex 
in the simulations. The standard value $-0.7$~dex is usually 
adopted, following in situ spectroscopic determination from 
\citet{1995AJ....109.1095G} and photometric star count 
determinations \citep{Robin1996A&A...305..125R,1999A&A...348...98B}. 
The low metallicity tail of the thick disc seems to represent
a weak contribution to general star counts \citep{1993AJ....105..539M}.
It was neglected here.
An internal metallicity dispersion among the thick disc 
population is allowed. The standard value for this dispersion is 0.25~dex with 
no metallicity gradient.
 
The thick disc density law is assumed to be a truncated exponential: at 
large distances the law is exponential, at short distances it is a parabola 
\citep{Robin1996A&A...305..125R}. This formula, given in equation~\ref{eq1},
ensures the continuity and derivability of the density law (contrary to a 
true exponential) and eases the computation of the potential. 

\begin{equation}
\label{eq1}
\rho \propto \left\{\begin{array}{ll}
\exp{(-\frac{R-R_{\small \sun}}{h_{R}})}*(1-\frac{1/h_{z}}{x_l*(2.+x_l/h_{z})}*z^{2}) & 
\mbox{ if } z\leq x_l \\
\exp{(-\frac{R-R_{\small \sun}}{h_{R}})}*\exp({-\frac{z}{h_{z}})} &
\mbox{ if } z> x_l \\
\end{array}
\right.
\end{equation}

Three parameters define the density
along the z axis : $h_{z}$, the scale height, $\rho_{0}$ the local
density and $x_l$ the distance above the plane where the density law
becomes exponential. This third parameter is fixed by continuity of
$\rho(z)$ and its derivative. It varies
with the choice of scale height and local density following the
potential.

\subsection{The spheroid}

We assume a homogeneous population of spheroid stars with a
short period of star formation. We thus use the \citet{Bergbush92} oxygen 
enhanced models, assuming an age of 14 Gyr (until more constraints on the age 
are available), a mean
metallicity of $-1.7$~dex and a dispersion of 0.25 about this value. No 
galactocentric
gradient is assumed. The colours are obtained from model atmospheres of
\citet{Lejeune1997A&AS..125..229L,Lejeune1998A&AS..130...65L}

The density law and IMF slope is the one determined from deep star counts 
in numerous directions, as described in \citet{Robin2000A&A}. This is a power 
law dn/dm~$\propto$~m$^{-1.9}$, an axis ratio of 0.7 and a local density of 
1.64~10$^{-4}$~stars~pc$^{-3}$, excluding the white dwarfs.

\section{Data analysis method}
\label{sec4}

Population synthesis simulations have been computed in each observed field
using photometric errors as close as possible to the true observational
errors, generally growing as a function of the
magnitude and assumed to be Gaussian. Monte Carlo simulations were done 
in a solid angle larger or equal to the data in order to minimize the 
Poisson noise.

We then compared the number of stars produced by the model 
with the observations in the selected region of the plane (magnitude, colour)
and computed the likelihood that the observed data fits the 
model \citep[following the method described in][appendix C]{Bienayme1987a}.

\begin{figure*}
\centering
\includegraphics{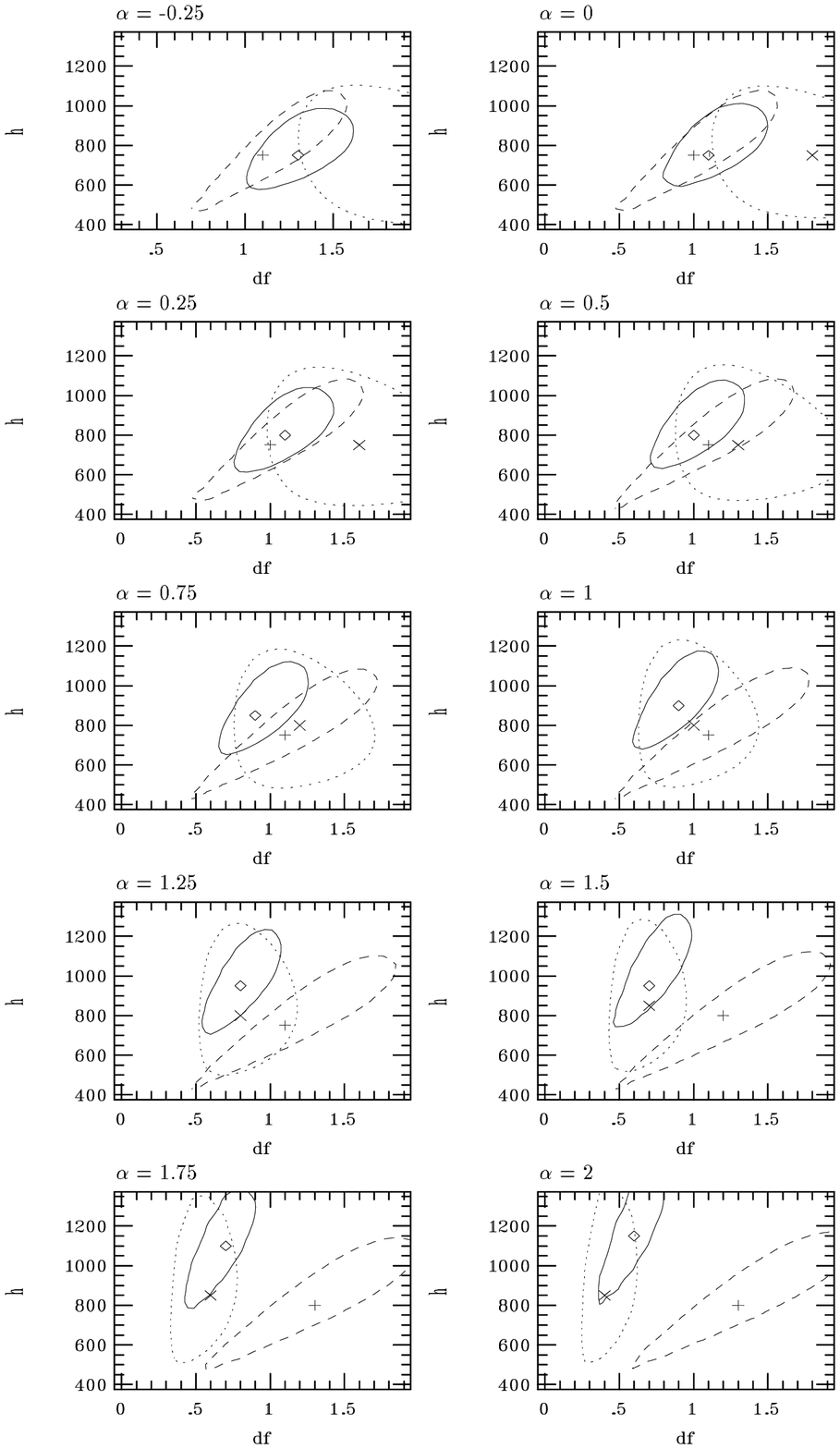}
   \caption{Iso-contour likelihoods at 1 $\sigma$ as a function of scale 
height h and density df for different IMF slopes $\alpha$. Plus and dashed 
line: DENIS fields. Diamond and solid line: deep fields (V $\geq$ 20). Cross
and dotted line: SA57.}
   \label{fig2}
\end{figure*}

The likelihood has been computed for a set of models, with varying thick disc
parameters : scale height range 400 to 1400 pc, scale length range 2 to 4 kpc
and the IMF slope $\alpha$ from -0.25 to 2.
In place of the local density we used a new parameter to try to overcome
the degeneracy between the scale height and the local density. Since the
number of stars is expected to vary as the volume times the scale height
times the local density, we used the following density parameter, df :
\[ \mbox{df} = \mbox{local density} \times \mbox{scale height}^2 \]
While this parameter does not reduce the degeneracy in shallow star counts
(magnitude smaller than 20), it breaks the denegeracy in deeper counts, as can be
seen in the likelihood contours in figure~\ref{fig2}. The df parameter varies 
from 0 to 2. 

The confidence limits of the estimated parameters are determined by the            
likelihood level which can be reached by random changing
of the sample:                       
a series of simulated random samples are produced using the set of model
parameters. 
The rms dispersion of the likelihood about the mean of this series gives 
an estimate of the likelihood fluctuations due to the random noise. It 
is then used to compute the confidence limit. Resulting 
errors are not strictly 
speaking standard errors; they give only an order of magnitude. 


\section{Results}
\label{sec5}
\subsection{Luminosity and mass function}

The IMF slope is best constrained when separately studying 
the fields, as the data do not cover the same mass range of thick disc stars.
Shallow and deep fields give constraints on different parts of the luminosity
function.
Thick disc stars in DENIS fields have masses greater than 0.6\Msun. The mass 
range of thick disc stars in deep counts is 0.2 to 0.6\Msun. The 
deepest field towards the North Galactic Pole (SA57) is dominated by stars 
with masses between 0.2 and 0.4\Msun. Figure~\ref{fig2} shows 
iso-contour likelihoods as a function
of scale height h and density df for different IMF slopes, for DENIS fields, 
deep fields, and SA57, separately. An IMF slope $\alpha \geq$~1.25 does not 
allow an
acceptable solution for all the fields. However, a lower IMF slope, 
$\alpha$~=~0.5, gives an agreement for all three magnitude intervals.

\begin{figure}
\centering
\includegraphics[width=7cm]{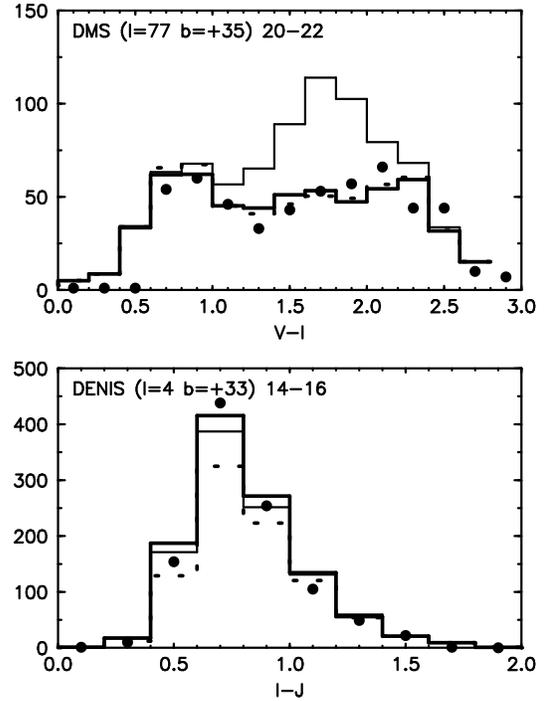}
   \caption{Colour distributions of a DMS field in the magnitude bin I=20-22
and a DENIS field in the magnitude bin I=14-16. The dots show the observations.
Thick lines are predicted number of stars by the model assuming $\alpha$ = 
0.5, h = 800 pc and df = 1. Thin lines: $\alpha$ = 2, h = 800 pc, df = 1. 
Dotted lines: $\alpha$ = 1.7, h = 1100 pc, df = 0.7.}
   \label{fig3}
\end{figure}

Unlike DENIS fields, the best solution for deep fields is very 
sensitive to the IMF slope. This is also shown in figure~\ref{fig3} (solid 
lines). In a recent paper
\citep{Robin2000A&A}, we found an IMF slope $\alpha$~=~1.7 for the thick 
disc from analysis of a set of deep fields only. It appears now that a 
degeneracy between the parameters was not overcome and that the choice of
a slightly larger scale height led to a higher IMF slope (dotted
lines in figure~\ref{fig3}). However, we 
lacked a wide enough mass range to put a strong constraint on it.
The solution given in the previous paper is not viable when we compare it to
the DENIS star counts (dotted lines in figure~\ref{fig3}). 
The only acceptable solution for the combination
of deep and shallow star counts is a smaller IMF slope.

We also tried to combine several IMF slopes in different mass intervals.
The likelihood value is slightly better when $\alpha$~=~2 for DENIS fields
and $\alpha$~=~1 for deep fields (see table~2). We
tested an IMF with a change of slope at 0.6\Msun (M$_V$~=~8), m$^{-2}$ at 
higher masses and m$^{-1}$ at lower masses, but the global likelihood value is 
smaller than when considering a single power law m$^{-0.5}$ (table~2) and the
best solutions for the DENIS, deep, and Sa57 fields are not in agreement.

\citet{Paresce2000ApJ} emphasize that the IMF of globular clusters of
similar abundances, such as 47 Tuc, have the shape of a lognormal
distribution: it rises as m$^{-1.6}$ in the range 0.3-0.8\Msun, then drops as 
m$^{-0.2}$ below 0.3\Msun. We find that such a lognormal 
distribution does not give a single solution acceptable for the DENIS, deep, 
and Sa57 fields (see also table~2 for the likelihood values).

The fainter thick disc stars in the deep fields are in the mass range 
0.2-0.5\Msun. The deepest bin 22-24 in Sa57 contains thick 
disc stars with masses from 0.1 to 0.4\Msun. This field alone does not give 
enough constraints to determine if an IMF with a change of slope around 
0.3\Msun would give a better agreement because of the sample size. Large scale 
surveys at this depth, like MEGACAM or VISTA projects,
would be necessary to definitely choose between several power laws or a 
lognormal IMF.

\subsection{Scale height and scale length}

\begin{figure}
\centering
\includegraphics{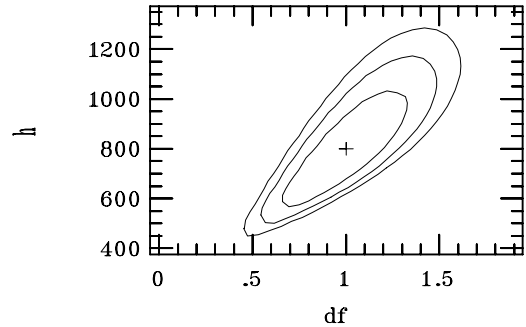}
   \caption{Iso-contour likelihoods at 1, 2, and 3 $\sigma$  as a function of 
scale height h and density df for all the fields together, considering an IMF
slope $\alpha$ = 0.5.}
   \label{fig4}
\end{figure}

\begin{figure*}
\centering
\includegraphics{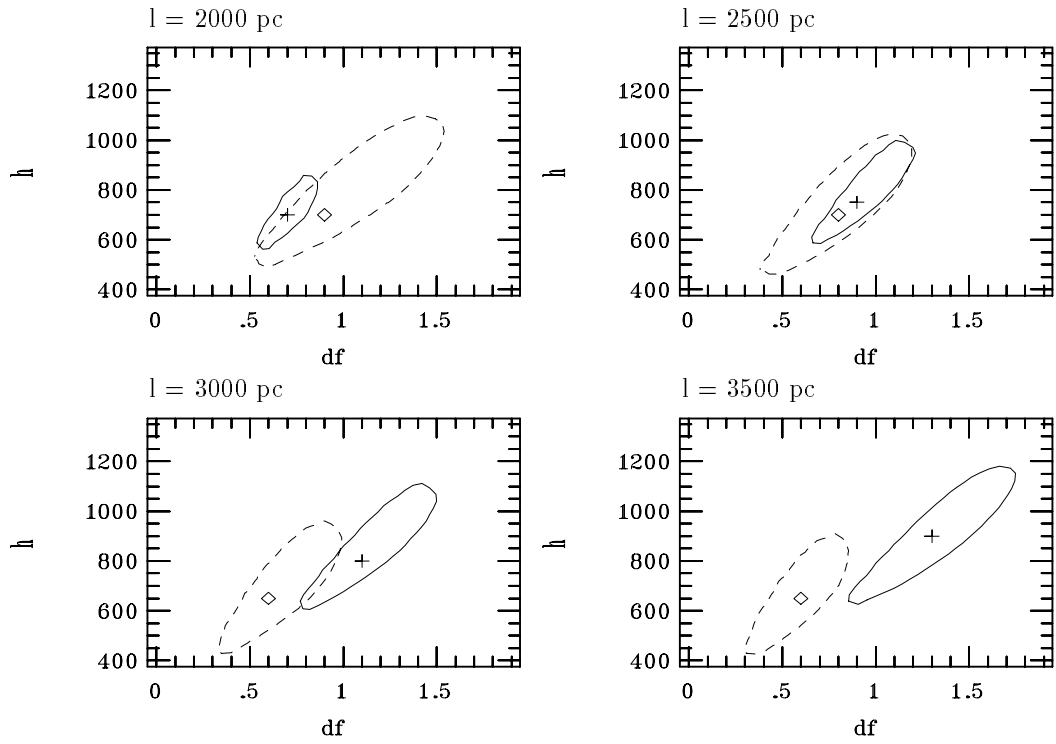}
   \caption{Iso-contour likelihoods at 1 $\sigma$ as a function of scale 
height h and density df for different scale lengths l. Plus and solid line: 
fields towards the galactic centre. Diamond and dashed line: fields in the 
anticentre direction.}
   \label{fig5}
\end{figure*}

As shown in figure~\ref{fig4}, our best fit model for all the fields together 
gives a scale height of 800 pc with df = 1. Iso-contours at 1, 2 and 3 
$\sigma$ are also plotted. Assuming this scale height and density, the maximum 
likelihood is obtained for a scale length of 2500 pc, but smaller or greater 
values are still acceptable. We now consider separately the fields in the 
anticentre (135$^\circ$ $\leq$ l $\leq$ 225$^\circ$) and the center (l $\geq$ 
315$^\circ$ or l$\leq$ 45$^\circ$), at medium galactic latitude ($|$b$|$ 
$\leq$ 50$^\circ$) where the effects 
of the scale length are most important. Iso-contours at 1 $\sigma$ are 
plotted in figure~\ref{fig5} with solid lines for the centre fields and dashed lines 
for the anticentre fields, the scale length ranging from 2000 pc to 3500 pc. 
A scale length of 2500 pc gives the best agreement between the centre and 
anticentre fields. Disc scale length is about 2500 pc, from the most recent studies
\citep{Robin1992A&A...265...32R,Fux94,Ruphy1996A&A...313L..21R}. 
The thick disc one seems to be very similar. The lack of deeper fields close 
to the anticentre does not allow us to better constrain this parameter or to 
reveal the presence of a flare.

\subsection{Sensitivity to other parameters}

\subsubsection{The Age of the thick disc}
Considering a younger thick disc of 11 Gyr instead of 14 Gyr gives the same 
results over the scale height, density and scale length. The IMF slope that 
allows us to reconcile deep and shallow fields is slightly different, 
$\alpha$~=~0.75, but still within our error bars. However, the absolute likelihoods 
are not as good as the ones obtained 
with an age of 14 Gyr. This parameter should mainly be determined from the 
turnoff position with accurate enough counts in a homogeneous system. The color
shift in I-J for a 8 Gyr to a 14 Gyr thick disc is only of 0.06 magnitude at 
the turnoff, out of reach at the DENIS precision.

\subsubsection{The spheroid IMF}
We have considered an IMF slope $\alpha$~=~1.9 for the spheroid, as derived by
\citet{Robin2000A&A} from the deep star counts. Even considering an IMF 
slope as low as $\alpha$ = 0.75 for the spheroid \citep{1998ApJ...503..798G}
does not change our results concerning the thick disc IMF slope. The best fit
parameters are h~=~850~pc and df~=~1.1, well within our error bars.

\subsubsection{Binary effects}
\begin{figure}
\centering
\includegraphics[angle=-90,width=7cm]{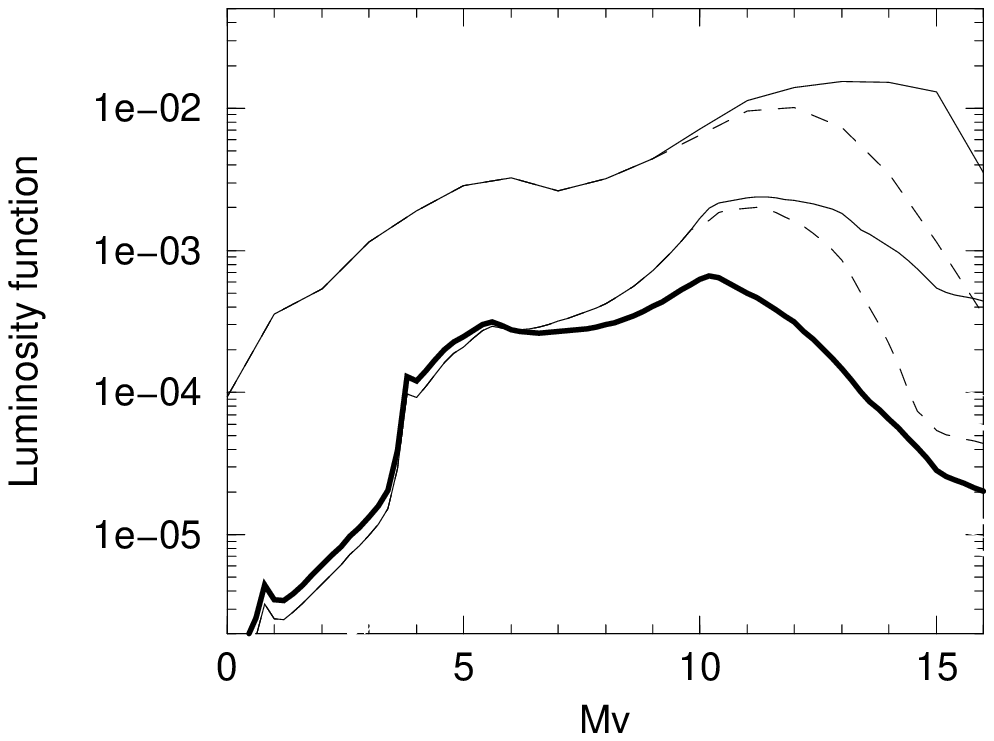}
   \caption{Upper curves: luminosity function in stars pc$^{-3}$ mag$^{-1}$ of the 
thin disc with an IMF 
slope $\alpha$ = 1.6. Solid line: single star luminosity function. Dashed 
line: system luminosity function. Lower curves: luminosity function of the 
thick disc. Solid thin line: single star luminosity function with $\alpha$ = 
2. Dashed line: system luminosity function with $\alpha$ = 2. Thick line: 
single star luminosity function with $\alpha$ = 0.5.}
   \label{fig6}
\end{figure}

The thin disc luminosity function that we used is a nearby luminosity function
that does not take into account the fact that we may observe systems instead
of single stars. 
\citet{Kroupa2000ASP} showed that the difference between the nearby 
luminosity function and the system one comes from the binary fraction, nearby
systems being computed as true single stars, while systems are not resolved
in remote star counts. Hence, a correction is to be applied to the nearby
luminosity function to take into account this effect in the counts.
Following Kroupa, we applied a correction to the luminosity function
used by the model in order to match the system luminosity function, as shown
in figure~\ref{fig6} (upper curves).
Only the deepest fields could be sensitive to this effect. When we apply this 
correction, the best fit thick disc parameters remain unchanged.

For the thick disc, the luminosity function considered in the model is the
luminosity function of single stars. Knowing little about the binarity in the
thick disc, we temporarily applied the same correction as for the thin 
disc on the
luminosity function (see figure~\ref{fig6}, lower curves). 
The best fit is obtained for the same density parameters h and 
df, but it slightly displaces the best IMF slope to $\alpha$~=~0.75. 
The likelihood values are slightly higher than those obtained with a single 
luminosity function (table~2).

\section{Discussions and conclusions}
\label{sec6}
We have estimated the thick disc density law parameters and mass function
by using a wide set of data at high and intermediate galactic latitudes, in 
the visible and the near infrared. The best fit model has a scale height of
800 pc, a scale length of 2500 pc and a density of 10$^{-3}$ stars pc$^{-3}$ 
or 7.1 10$^{-4}$ \Msun pc$^{-3}$ for Mv $\leq$ 8, that is 6.2\% of the thin 
disc density. This result confirms the values obtained in 1996 from a smaller 
number of fields and mass ranges \citep{Robin1996A&A...305..125R}. 

For the first time, we determined the thick disc mass function over a large
mass range by the study of shallow star counts dominated by stars at 
the turnoff combined with deep star counts. The IMF of the thick disc seems to 
follow a power law dn/dm~$\propto$~m$^{-0.5}$ in the mass range 0.2-0.8\Msun. We found 
no evidence of a change of slope at lower masses, but we only have one field deep 
enough to constrain the IMF at low masses.

The only point of comparison for the thick disc mass function is the globular
clusters of similar metallicities, where the conditions of star formation
could be comparable. However, clusters are subject to bias
with regard to the field because of possible mass segregation effects. From seven
clusters, \citet{1999A&A...345..485P} found a mean $\alpha$ of 0.89 at masses
below 0.6\Msun but the range covers 1.22 to 0.53, which is compatible
with our measurements in the field.

\citet{Paresce2000ApJ} found for similar clusters an indication 
of a lognormal IMF rather
than a power law. If fitted by a power law in the mass range 
0.3-0.8\Msun, their IMF should be approximated by $\alpha$=1.6, but going down to 
 $\alpha$=0.2 at m$<$0.3 \Msun.
The result clearly depends on the way an IMF, which globally is not a power 
law,
is fitted in different mass intervals by portions of power laws.
While our result seems not to favor a lognormal IMF compared to a 
power law slope in the mass range 0.2-0.8\Msun, more data at lower masses may 
change our conclusion in the future.

Values of the spheroid IMF slope in the field range between 1.7$\pm$0.3 from 
a small local sample \citep{1997A&A...328...83C}, 0.75 from HST star
counts \citep{1998ApJ...503..798G}, and the higher value $\alpha$=1.9 determined from
wide field star counts \citep{Robin2000A&A}. However, the latter is valid 
at masses m$>$0.3\Msun 
while the former go deeper to masses of about 0.1\Msun and are less sensitive
to higher masses because of the small size of the samples. A change of 
slope between
intermediate masses and low masses, as found in the disc, may explain 
the apparent discrepancy, which is also partly due to the 
Poisson noise in these relatively small samples.

\citet{1998simf.conf..201S} suggests it is not valid to rely upon a mean IMF, 
since variations are 
too high from one measurement to the other at the present time, but he 
identifies no tendency correlated 
with any physical parameter. On the other hand, \citet{Kroupa2000AGM} finds
that young clusters seem to have a steeper slope for m$<1$\Msun and ancient
globular clusters have $\alpha>0$ but closer to 0 than the field. He concludes
that star formation produces relatively more low mass stars at later galactic
epochs. His galactic field IMF for the disc has a slope
of 1.3$\pm$0.5 at 0.1\Msun$<$m$<$0.5\Msun and 2.3$\pm$0.3 at 
0.5\Msun$<$m$<$1\Msun. 

\begin{figure*}
\centering
\includegraphics[angle=-90,width=10cm]{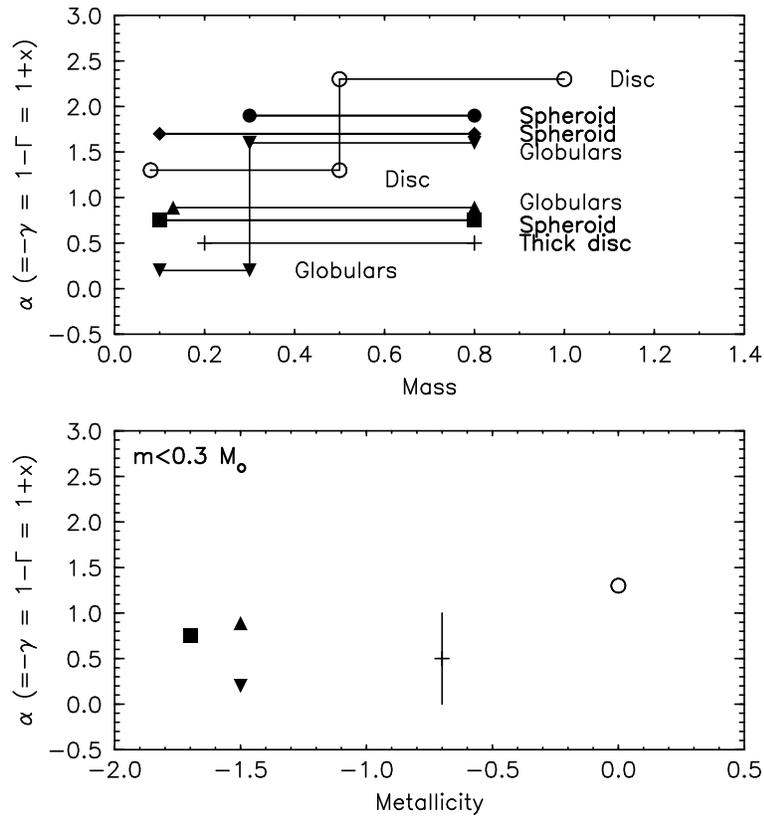}
   \caption{IMF slope $\alpha$ versus mass in \Msun and metallicity. $\gamma$,
$\Gamma$ and x are also power-law indices that may be used by other authors.
Thin lines and filled symbols: globular clusters (triangles down: 
\citet{Paresce2000ApJ}, triangles up: \citet{1999A&A...345..485P}). Thick lines 
and filled symbols: spheroid field stars (circles: \citet{Robin2000A&A}, 
diamonds: \citet{1997A&A...328...83C}, squares: \citet{1998ApJ...503..798G}). 
plus: thick disc stars. Open circles: disc stars \citep{2001MNRAS.322..231K}.}
   \label{fig7}
\end{figure*}

If we consider the alpha-plot (figure 14 of \citet{Kroupa2000AGM}), our 
measurement of a thick disc mass function similar to globular clusters seems 
to corroborate Kroupa's conclusion. The thick disc mass function 
at low masses seems to be flatter than the thin disc and comparable with the 
spheroid. Figure~\ref{fig7} shows the 
alpha-plot versus metallicity for low-mass stars (m$<$3\Msun), for which the
effect is clearer.
However, when taking into account other halo IMFs measured in the field, like
\citet{1997A&A...328...83C}, the correlation
is less clear because of larger error bars due to the small size of the
sample, as seen in the alpha-plot versus mass (figure~\ref{fig7}). Clearly, 
the answer depends on
a better determination of the IMF in the field, which can be done by exploring
a larger mass range (going deeper) and having larger samples. This is
promising in that several large scale surveys are on the way or planned for 
the near future.

\begin{acknowledgements}
The authors thank J\'er\^ome Bouvier for giving them access to his data, S\'ebastien
Derriere and Bertrand Bassang who helped on the exploitation of the DENIS database,
the whole DENIS staff and all the DENIS observers who collected the data. 
The DENIS project is supported by the SCIENCE and the Human Capital and Mobility plans 
of the European Commission under grants CT920791 and CT940627 in France, by l'Institut 
National des Sciences de l'Univers, the Ministère de l'Éducation Nationale and the Centre 
National de la Recherche Scientifique (CNRS) in France, by the State of Baden-Württemberg 
in Germany, by the DGICYT in Spain, by the Sterrewacht Leiden in Holland, by the
Consiglio Nazionale delle Ricerche (CNR) in Italy, by the Fonds zur Förderung der 
wissenschaftlichen Forschung and Bundesministerium für Wissenschaft und Forschung in 
Austria, and by the ESO C \& EE grant A-04-046.
\end{acknowledgements}

\begin{table*}
\label{table1}
\caption{Characteristics of the selected fields.}
\begin{tabular}{lllll}
\hline\\
Reference	&Field	&Area	&Bands	&Magnitude\\
		&coordinates	&(deg$^2$)	&	&range\\
\hline\\
Ojha et al. (1994a,b, 1996)	&l=4$^\circ$, b=+47$^\circ$	&15.5	&B,V	&V=15-17\\
			&l=278$^\circ$, b=+47$^\circ$	&20.8	&B,V	&V=16-18.5\\
			&l=168$^\circ$, b=+48$^\circ$	&7.13	&B,V	&V=15-18\\
Chiu (1980)		&l=189$^\circ$, b=+21$^\circ$	&0.10	&B,V	&V=18-20\\
Yee et al. (2000)	&l=166$^\circ$, b=-55$^\circ$	&0.39	&V,I	&V=17-20\\
Borra \& Lepage (1986)	&l=197$^\circ$, b=+38$^\circ$	&0.25	&B,V	&V=20-22\\
Reid \& Majewski (1993)	&North Galactic Pole		&0.30	&B,V	&V=19-22\\
DMS			&l=129$^\circ$, b=-63$^\circ$	&0.14	&V,I	&V=18-22\\
			&l=248$^\circ$, b=+47$^\circ$	&0.08	&V,I	&V=18-22\\
			&l=337$^\circ$, b=+57$^\circ$	&0.16	&V,I	&V=18-22\\
			&l=77$^\circ$, b=+35$^\circ$	&0.15	&V,I	&V=18-22\\
			&l=52$^\circ$, b=-39$^\circ$	&0.15	&V,I	&V=18-22\\
			&l=68$^\circ$, b=-51$^\circ$	&0.15	&V,I	&V=18-22\\
Bouvier et al. (1998)	&l=117$^\circ$, b=-23$^\circ$	&1.48	&I,R	&R=17-22.4\\
			&l=206$^\circ$, b=+32$^\circ$	&0.55	&I,R	&R=17-22\\
Sa57			&l=69$^\circ$, b=+85$^\circ$	&0.16	&V,I	&V=20-24\\
\hline
\end{tabular}
\end{table*}

\begin{table*}
\label{table2}
\caption{Likelihood values obtained for the best fit parameters h and df when 
considering different thick disc IMF. The best fit parameters h and df are 
shown in figure~\ref{fig2} for the DENIS, 
deep and Sa57 fields in the case of an IMF following a single power law.}
\begin{tabular}{lllll}
\hline\\
dn/dm		&DENIS fields	&deep field	&SA57	        &all fields\\
\hline\\
m$^{-0.5}$	&-1830$\pm$200	&-2500$\pm$150	&-92$\pm$10	&-5300$\pm$400\\
m$^{-0.75}$	&-1790		&-2420		&-88		&-5250\\
m$^{-1}$	&-1760		&-2370		&-85		&-5370\\
m$^{-1.5}$	&-1730		&-2370		&-79		&-6260\\
m$^{-2}$	&-1730		&-2480		&-77		&-8130\\
m$^{-2}$ (m$>$0.6\Msun), m$^{-1}$ (m$<$0.6\Msun) 
		&-1730		&-2570		&-85		&-7230\\
$\frac{1}{\mbox{m}}$exp-$\frac{[\mbox{log(m}/0.34)]^2}{2(0.4)^2}$ (lognormal) 
		&-1710		&-2510		&-89		&-6310\\
m$^{-0.75}$ + binary correction &-1820	&-2340		&-87	&-5100\\

\hline
\end{tabular}
\end{table*}

\begin{thebibliography}{40}
\expandafter\ifx\csname natexlab\endcsname\relax\def\natexlab#1{#1}\fi

\bibitem[{{Baraffe} {et~al.}(1998){Baraffe}, {Chabrier}, {Allard}, \&
  {Hauschildt}}]{1998A&A...337..403B}
{Baraffe}, I., {Chabrier}, G., {Allard}, F., \& {Hauschildt}, P.~H. 1998, A\&A,
  337, 403

\bibitem[{Bergbush \& VandenBerg(1992)}]{Bergbush92}
Bergbush, P.~A. \& VandenBerg, D.~A. 1992, ApJS, 81, 163

\bibitem[{{Bienaym\'e} {et~al.}(1987{\natexlab{a}}){Bienaym\'e}, {Robin}, \&
  {Cr\'ez\'e}}]{Bienayme1987a}
{Bienaym\'e}, O., {Robin}, A.~C., \& {Cr\'ez\'e}, M. 1987{\natexlab{a}}, A\&A,
  186, 359

\bibitem[{{Bienaym\'e} {et~al.}(1987{\natexlab{b}}){Bienaym\'e}, {Robin}, \&
  {Cr\'ez\'e}}]{Bienayme1987b}
{Bienaym\'e}, O., {Robin}, A.~C., \& {Cr\'ez\'e}, M. 1987{\natexlab{b}}, A\&A, 180, 
94

\bibitem[{Borra \& Lepage(1986)}]{Borra86}
Borra, E. \& Lepage, R. 1986, J, 92, 203

\bibitem[{{Bouvier} {et~al.}(1998){Bouvier}, {Stauffer}, {Martin}, {Barrado y
  Navascues}, {Wallace}, \& {Bejar}}]{Bouvier1998A&A}
{Bouvier}, J., {Stauffer}, J.~R., {Martin}, E.~L., et al. 1998, A\&A, 336, 490

\bibitem[{{Buser} {et~al.}(1999){Buser}, {Rong}, \&
  {Karaali}}]{1999A&A...348...98B}
{Buser}, R., {Rong}, J., \& {Karaali}, S. 1999, A\&A, 348, 98

\bibitem[Chabrier \& Mera(1997)]{1997A&A...328...83C} 
Chabrier, G. \& Mera, D. 1997, A\&A, 328, 83 

\bibitem[{Chiu(1980)}]{Chiu80b}
Chiu, L.-T. 1980, AJ, 85, 812.

\bibitem[{{Cr\'ez\'e} {et~al.}(1998){Cr\'ez\'e}, {Chereul}, {Bienaym\'e}, \&
  {Pichon}}]{Creze1998A&A}
{Cr\'ez\'e}, M., {Chereul}, E., {Bienaym\'e}, O., \& {Pichon}, C. 1998, A\&A, 329,
  920

\bibitem[{{Epchtein} {et~al.}(1997){Epchtein}, {De Batz}, {Capoani},
  {Chevallier}, {Copet}, {Fouque}, {Lacombe}, {Le Bertre}, {Pau}, {Rouan},
  {Ruphy}, {Simon}, {Tiphene}, {Burton}, {Bertin}, {Deul}, {Habing},
  {Borsenberger}, {Dennefeld}, {Guglielmo}, {Loup}, {Mamon}, {Ng}, {Omont},
  {Provost}, {Renault}, {Tanguy}, {Kimeswenger}, {Kienel}, {Garzon}, {Persi},
  {Ferrari-Toniolo}, {Robin}, {Paturel}, {Vauglin}, {Forveille}, {Delfosse},
  {Hron}, {Schultheis}, {Appenzeller}, {Wagner}, {Balazs}, {Holl}, {Lepine},
  {Boscolo}, {Picazzio}, {Duc}, \& {Mennessier}}]{Epchtein1997Msngr..87...27E}
{Epchtein}, N., {De Batz}, B., {Capoani}, et al. 1997, The Messenger, 87, 27

\bibitem[{{Epchtein} {et~al.}(1999){Epchtein}, {Deul}, {Derriere},
  {Borsenberger}, {Egret}, {Simon}, {Alard}, {Bal{\'a}zs}, {de Batz}, {Cioni},
  {Copet}, {Dennefeld}, {Forveille}, {Fouqu{\'e}}, {Garz{\'o}n}, {Habing},
  {Holl}, {Hron}, {Kimeswenger}, {Lacombe}, {Le Bertre}, {Loup}, {Mamon},
  {Omont}, {Paturel}, {Persi}, {Robin}, {Rouan}, {Tiph{\`e}ne}, {Vauglin}, \&
  {Wagner}}]{1999A&A...349..236E}
{Epchtein}, N., {Deul}, E., {Derriere}, S., et al. 1999, A\&A, 349, 236

\bibitem[{Fux \& Martinet(1994)}]{Fux94}
Fux, R. \& Martinet, L. 1994, A\&A, 287, L21

\bibitem[{{Gilmore} {et~al.}(1995){Gilmore}, {Wyse}, \&
  {Jones}}]{1995AJ....109.1095G}
{Gilmore}, G., {Wyse}, R. F.~G., \& {Jones}, J.~B. 1995, AJ, 109, 1095

\bibitem[{{Gomez} {et~al.}(1997){Gomez}, {Grenier}, {Udry}, {Haywood},
  {Meillon}, {Sabas}, {Sellier}, \& {Morin}}]{Gomez1997ESASP}
{Gomez}, A.~E., {Grenier}, S., {Udry}, S., et al. 1997, ESA SP-402: Hipparcos -
  Venice '97, 402, 621

\bibitem[Gould {et~al.}(1998)]{1998ApJ...503..798G} 
Gould, A., Flynn, C., \& Bahcall, J.~N. 1998, ApJ, 503, 798 

\bibitem[{{Gratton} {et~al.}(2000){Gratton}, {Carretta}, {Matteucci}, \&
  {Sneden}}]{Gratton2000A&A}
{Gratton}, R.~G., {Carretta}, E., {Matteucci}, F., \& {Sneden}, C. 2000, A\&A,
  358, 671

\bibitem[{{Hall} {et~al.}(1996){Hall}, {Osmer}, {Green}, {Porter}, \&
  {Warren}}]{Hall1996ApJS}
{Hall}, P.~B., {Osmer}, P.~S., {Green}, R.~F., {Porter}, A.~C., \& {Warren},
  S.~J. 1996, ApJS, 104, 185

\bibitem[{{Haywood} {et~al.}(1997){Haywood}, {Robin}, \&
  {Cr\'ez\'e}}]{Haywood1997A&A...320..440H}
{Haywood}, M., {Robin}, A.~C., \& {Cr\'ez\'e}, M. 1997, A\&A, 320, 440

\bibitem[{{Jahreiss} \& {Wielen}(1997)}]{Jahreiss1997ESASP}
{Jahreiss}, H. \& {Wielen}, R. 1997, ESA SP-402: Hipparcos - Venice '97, 402,
  675

\bibitem[{{Kroupa}(2000a)}]{Kroupa2000AGM}
{Kroupa}, P. 2000a, PASP, 228, 187

\bibitem[{{Kroupa}(2000b)}]{Kroupa2000ASP}
{Kroupa}, P. 2000b, in ASP Conf. Ser. : Dynamics of Star Clusters and the
  Milky Way, 201

\bibitem[{{Kroupa}(2001)}]{2001MNRAS.322..231K}
{Kroupa}, P. 2001, MNRAS, 322, 231

\bibitem[{{Lejeune} {et~al.}(1997){Lejeune}, {Cuisinier}, \&
  {Buser}}]{Lejeune1997A&AS..125..229L}
{Lejeune}, T., {Cuisinier}, F., \& {Buser}, R. 1997, A\&AS, 125, 229

\bibitem[{{Lejeune} {et~al.}(1998){Lejeune}, {Cuisinier}, \&
  {Buser}}]{Lejeune1998A&AS..130...65L}
{Lejeune}, T., {Cuisinier}, F., \& {Buser}, R. 1998, A\&AS, 130, 65

\bibitem[{Mera {et~al.}(1996)Mera, Chabrier, \& Baraffe}]{Mera96b}
Mera, D., Chabrier, G., \& Baraffe, I. 1996, ApJ, 459, L87

\bibitem[{{Morrison}(1993)}]{1993AJ....105..539M}
{Morrison}, H.~L. 1993, AJ, 105, 539

\bibitem[{{Ojha} {et~al.}(1996){Ojha}, {Bienaym\'e}, {Robin}, {Cr\'ez\'e}, \&
  {Mohan}}]{Ojha1996}
{Ojha}, D.~K., {Bienaym\'e}, O., {Robin}, A.~C., {Cr\'ez\'e}, M., \& {Mohan},
  V. 1996, A\&A, 311, 456

\bibitem[{{Ojha} {et~al.}(1999){Ojha}, {Bienaym{\'e}}, {Mohan}, \&
  {Robin}}]{1999A&A...351..945O}
{Ojha}, D.~K., {Bienaym{\'e}}, O., {Mohan}, V., \& {Robin}, A.~C. 1999, A\&A,
  351, 945

\bibitem[{{Osmer} {et~al.}(1998){Osmer}, {Kennefick}, {Hall}, \&
  {Green}}]{Osmer1998ApJS}
{Osmer}, P.~S., {Kennefick}, J.~D., {Hall}, P.~B., \& {Green}, R.~F. 1998,
  ApJS, 119, 189

\bibitem[{{Paresce} \& {De Marchi}(2000)}]{Paresce2000ApJ}
{Paresce}, F. \& {De Marchi}, G. 2000, ApJ, 534, 870

\bibitem[Piotto \& Zoccali(1999)]{1999A&A...345..485P} Piotto, G.\ \& 
Zoccali, M. 1999, A\&A 345, 485 

\bibitem[{{Reid} \& {Majewski}(1993)}]{Reid1993ApJ...409..635R}
{Reid}, N. \& {Majewski}, S.~R. 1993, ApJ, 409, 635

\bibitem[{{Robin} {et~al.}(1992){Robin}, {Cr\'ez\'e}, \&
  {Mohan}}]{Robin1992A&A...265...32R}
{Robin}, A.~C., {Cr\'ez\'e}, M., \& {Mohan}, V. 1992, A\&A, 265, 32

\bibitem[{{Robin} {et~al.}(1996){Robin}, {Haywood}, {Cr\'ez\'e}, {Ojha}, \&
  {Bienaym\'e}}]{Robin1996A&A...305..125R}
{Robin}, A.~C., {Haywood}, M., {Cr\'ez\'e}, M., {Ojha}, D.~K., \& {Bienaym\'e},
  O. 1996, A\&A, 305, 125

\bibitem[{{Robin} {et~al.}(2000){Robin}, {Reyl{\'e}}, \&
  {Cr{\'e}z{\'e}}}]{Robin2000A&A}
{Robin}, A.~C., {Reyl{\'e}}, C., \& {Cr{\'e}z{\'e}}, M. 2000, A\&A, 359, 103

\bibitem[{{Ruphy} {et~al.}(1996){Ruphy}, {Robin}, {Epchtein}, {Copet},
  {Bertin}, {Fouque}, \& {Guglielmo}}]{Ruphy1996A&A...313L..21R}
{Ruphy}, S., {Robin}, A.~C., {Epchtein}, N., et al. 1996, A\&A, 313, L21

\bibitem[{{Scalo}(1998)}]{1998simf.conf..201S}
{Scalo}, J. 1998, in ASP Conf. Ser. 142: The Stellar Initial Mass Function
  (38th Herstmonceux Conference), 201

\bibitem[{{Sommer-Larsen} \& {Antonuccio-Delogu}(1993)}]{1993MNRAS.262..350S}
{Sommer-Larsen}, J. \& {Antonuccio-Delogu}, V. 1993, MNRAS, 262, 350

\bibitem[{Twarog(1980)}]{Twarog80}
Twarog, B. 1980, ApJS, 44, 1

\bibitem[{{Yee} {et~al.}(2000){Yee}, {Morris}, {Lin}, {Carlberg}, {Hall},
  {Sawicki}, {Patton}, {Wirth}, {Ellingson}, \& {Shepherd}}]{Yee2000ApJS}
{Yee}, H. K.~C., {Morris}, S.~L., {Lin}, H., et al. 2000, ApJS, 129, 475

\end{thebibliography}
\end{document}